\documentclass[aps,superscriptaddress,twocolumn,amssymb]{revtex4-1}
\usepackage{graphicx}
\DeclareMathSymbol{@}{\mathord}{letters}{"3B} 

\draft 

\begin{document}


\title{3D microwave cavity with magnetic flux control and enhanced quality factor} 


\author{Yarema Reshitnyk}
\affiliation{School of Mathematics and Physics, University of Queensland, Brisbane, Queensland 4072, Australia}
\author{Markus Jerger}
\affiliation{ARC Centre of Excellence for Engineered Quantum Systems, The University of Queensland, St Lucia QLD 4072
	Australia}
\author{Arkady Fedorov}
\email{a.fedorov@uq.edu.au}
\affiliation{School of Mathematics and Physics, University of Queensland, Brisbane, Queensland 4072, Australia}
\affiliation{ARC Centre of Excellence for Engineered Quantum Systems, The University of Queensland, St Lucia QLD 4072
	Australia}



\date{\today}

\begin{abstract}
Three-dimensional (3D) superconducting microwave cavities with large mode volumes typically have high quality factors ($>10^6$). This is due to a reduced sensitivity to surface dielectric losses, which is the limiting source of dissipation in two-dimensional transmission line resonators. In recent years, 3D microwave cavities have been extensively used for coupling and interacting with superconducting quantum bits (qubits), providing a versatile platform for quantum information processing and hybrid quantum systems. A current issue that has arisen is that 3D superconducting cavities do not permit magnetic field control of qubits embedded in these cavities. In contrast, microwave cavities made of normal metals can be transparent to magnetic fields, but experience a much lower quality factor ($\sim 10^4$), which negates many of the advantages of the 3D architecture. In an attempt to create a device that bridges a gap between these two types of cavities, having magnetic field control and high quality factor, we have created a hybrid 3D cavity. This new cavity is primarily composed of aluminium but also contains a small copper insert. We measured the internal quality factor of the hybrid cavity to be $102@000$, which is an order of magnitude improvement over all previously tested copper cavities. An added benefit to that our hybrid cavity possesses is that it also provides an improved thermal link to the sample that superconducting cavities alone cannot provide. In order to demonstrate precise magnetic control within the cavity, we performed spectroscopy of three superconducting qubits placed in the cavity, where individual control of each qubit's frequency was exerted with small wire coils attached to the cavity. A large improvement in quality factor and magnetic field control makes this 3D hybrid cavity an attractive new platform for circuit quantum electrodynamics experiments. 

\end{abstract}


\maketitle 

Three-dimensional (3D) superconducting cavities are well-known for their record high quality factors (Q's) of $\sim 10^9$ in the microwave frequency domain (see Ref.~\cite{Reagor2015} and references therein). It is even possible to achieve quality factors of several millions without any special requirements in manufacturing and on material purity. In recent years, 3D microwave cavities have been used for experiments with superconducting qubits, enabling a new architecture for superconducting quantum circuits~\cite{Paik2011}. Within this architecture the coherence of superconducting quantum circuits has drastically improved, which shed light on the origin of coherence limitations for superconducting qubits~\cite{Wang2015}. Currently, 3D microwave cavities with embedded superconducting qubits are one of the main platforms for circuit quantum electrodynamics experiments~\cite{Weber2014,Heeres2015} and are also being used in conjunction with other quantum systems~\cite{Probst2014,Tabuchi405}. Yet 3D superconducting cavities do come with a drawback.  The walls of superconductive cavities perfectly screen the interior of the cavity from external magnetic fields and thus controlling superconducting circuits with external magnetic flux is not possible. Additional difficulty arises from poor thermalization of the qubits as the superconductor of a cavity does not provide a good thermal link to the cold plate of a refrigerator. The most common way around of these issues is to use copper cavities, however, they have substantially smaller Q's, on the order of $10@000$~\cite{Rigetti2012}. More sophisticated measures to reduce losses and thermalize the cavity walls were also investigated~\cite{Bogorin2013}. 
Higher internal Q of the cavity for circuit-QED experiments with magnetically tunable qubits offers longer life-times of the qubits due to reduced Purcell limit. In addition decreasing internal loss channel while maintaining  strong coupling to the external circuitry is a necessary condition for quantum networking with microwave photons.

In this letter we present measurements completed in a 3D microwave cavity of rectangular geometry with dimensions 30.00$\,$mm$\times$4.60$\,$mm$\times$27.40$\,$mm. The majority of this cavity was machined out of standard aluminium (alloy 6061) with a small insert machined out of oxygen free copper (C10100) as shown in Fig.~1(b). The cavity has two microwave ports which can be used for either transmission or reflection measurements. Transmission measurements were used to perform spectroscopy and time-domain characterization of the superconducting qubits placed inside the cavity. The internal Q of the cavity was determined by measuring the bare cavity in reflection, with the second port disconnected. All room-temperature measurements were performed with a vector network analyzer (VNA). Low temperature measurements were done in a dilution refrigerator with use of additional filtering of the microwave lines, a low-noise HEMT amplifier to amplify the transmitted/reflected signal, and circulators to isolate the cavity from any noise from the amplifier and also to separate incoming and reflected signals when measuring in reflection (see Fig.~1(a)).

\begin{figure*}[htbp]
	\begin{center}
		\includegraphics[scale=0.75]{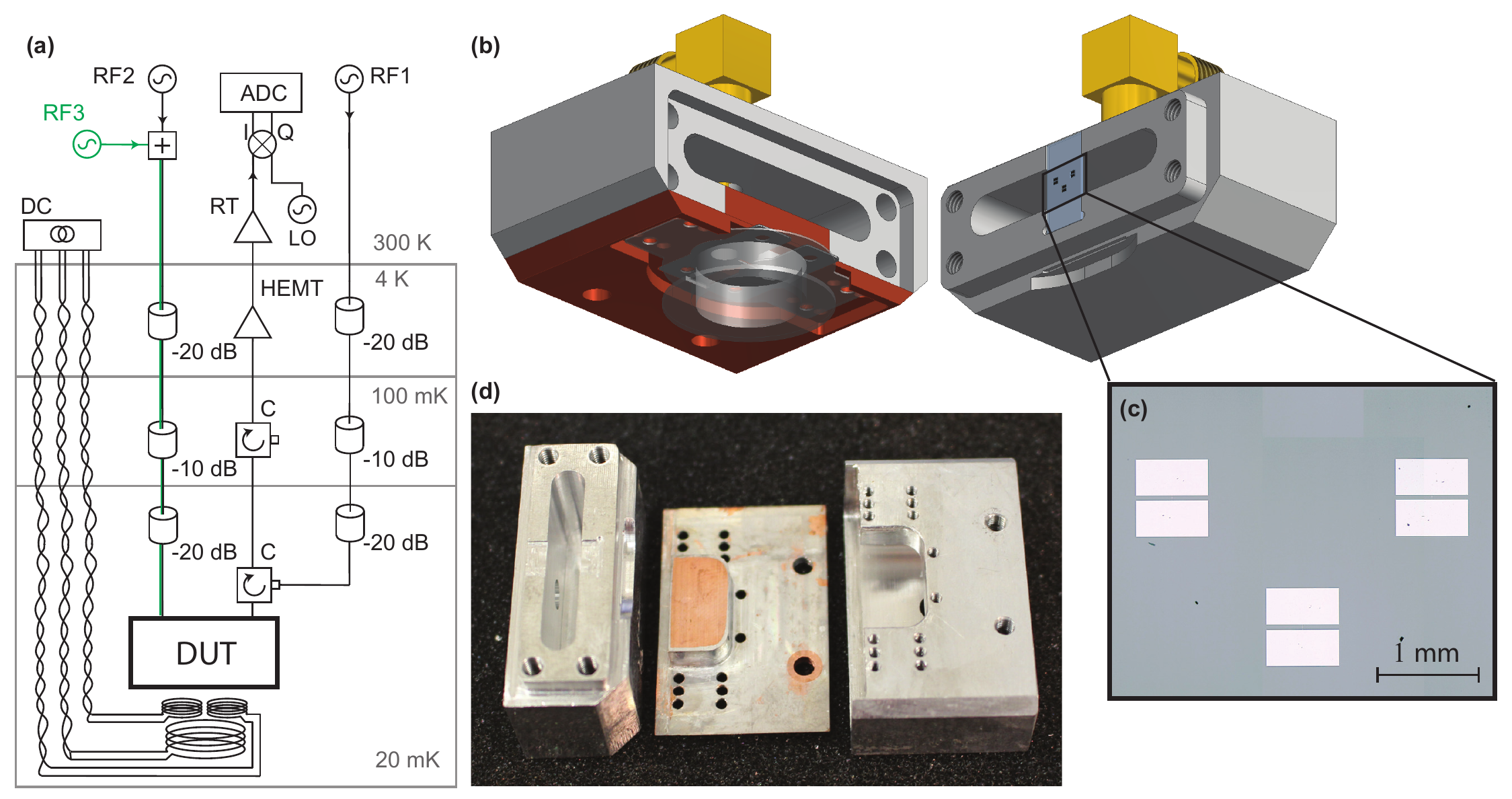}
		\caption{
			(a) Simplified diagram of the measurement setup at Millikelvin temperatures. To perform reflection measurements, a microwave tone (RF1) is applied to the output port of the cavity via a circulator (C). The reflected signal is amplified by a high-electron-mobility transistor (HEMT) amplifier at 4~K and a chain of room temperature (RT) amplifiers. The sample sits at 20~mK and is isolated from the higher temperature stages by an additional circulator in series. The amplified signal is down-converted to an intermediate frequency of 25~MHz in an IQ mixer, driven by a dedicated LO, and is digitized by an analogue-to-digital converter (ADC) for data analysis. To perform transmission measurement a microwave signal (RF2) is applied to the input port of the cavity. For spectroscopy of the qubits, an additional excitation microwave tone (RF3) is applied together with the measurement signal. The magnetic field inside the cavity is controlled by three coils attached to the exterior of the cavity and wired to individual current sources (DC) at room temperature.
			(b) 3D model of the cavity with a chip containing three flux tunable transmon qubits. Most of the cavity body is made of Al (gray) with a Cu insert (brown). The Cu insert allows magnetic field from the three mounted coils to penetrate the interior of the cavity for individual control of all qubits frequencies. In addition, the Cu insert provides a beneficial thermal link to the chip.
			(c) Magnified image of the three transmon qubits on the chip. Each qubit consists of two Al capacitor plates connected via a line interrupted by a micron size DC SQUID (not visible at this scale). 
			(d) Photograph of the cavity parts. To reduce losses due to possible gaps between Cu and Al parts, $\sim2\,\mu$m of Al were evaporated on the surfaces of the Cu insert which get in contact with Al.
		} 
		\label{fig1}
	\end{center}
\end{figure*}

We first performed a characterization of the cavity without superconducting qubits. The lowest resonant mode of the cavity (TE101) without a chip was found at $7.5905\,$GHz at a temperature of $\sim 20\,$mK. The quality factors of the cavities with similar geometries made only out of Cu were found to be $~4@800$ at room temperature and $11@000$ at Millikelvin temperatures. The quality factors of aluminium cavities made of the same alloy were found $>450@000$ (the measurement to determine the exact value of quality factor was not performed). As the cavities interior surface of the copper insert was designed to be normal to the electric field of the mode of interest, we expected Ohmic losses due to induced currents in the normal metal to be minimized, with Q of the hybrid cavity being substantially higher than one for the Cu cavity. 

To design the cavity we simulated our design using the eigenmode solver of Ansys HFSS. We first simulated the copper cavity at room temperature and obtained $Q\simeq18@000$.  With conductivity of copper at room temperature simulations matched the internal Q to the experimentally observed value of $4@800$. As it is hard to accurately predict the value of electrical conductivity of copper at Millikelvin temperatures we adjusted this parameter to match to value the Q of $11@000$ measured at $20\,$mK.  With the Millikelvin adjustments for copper, we performed a simulation of our hybrid design to find the quality factor of $180@000$ which is about more an order of magnitude improvement compared to the bare copper cavity. 

To experimentally determine the internal Q we measured in reflection from one port of the cavity, with the second port closed with Al tape. We adjusted the coupling of the port to obtain an external quality factor close to the value of the expected internal Q. Our first low temperature measurement of the hybrid cavity yielded $Q \sim 15@000$, only a moderate improvement compared to Cu cavity and nowhere near the simulated value. We attributed this behavior to small gaps between the superconducting and normal-metal interface of the cavity arising from the insert due to the manufacturing process (order of $\mu$m) and to roughness of the copper surface. Recent results also demonstrated that interfaces between different parts of the 3D cavities play a crucial role in defining their quality factors~\cite{Wang2015b}.

To better thermalize our qubit to the dilution refrigerator, the normal metal copper is in contact with the chip. The chip is placed between the two halves of the cavity, where one side of the chip faces the copper, and the other Al. While beneficial for thermalizing the qubit, this specific design can incur more losses due to an air gap between some parts of the cavity. We note that our HFSS simulations confirmed substantial decrease of the quality factor due to the gap, and surface roughness but we could not achieve quantitative agreement with the measured values. 

\begin{figure*}[htbp]
	\begin{center}
		\includegraphics[scale=0.75]{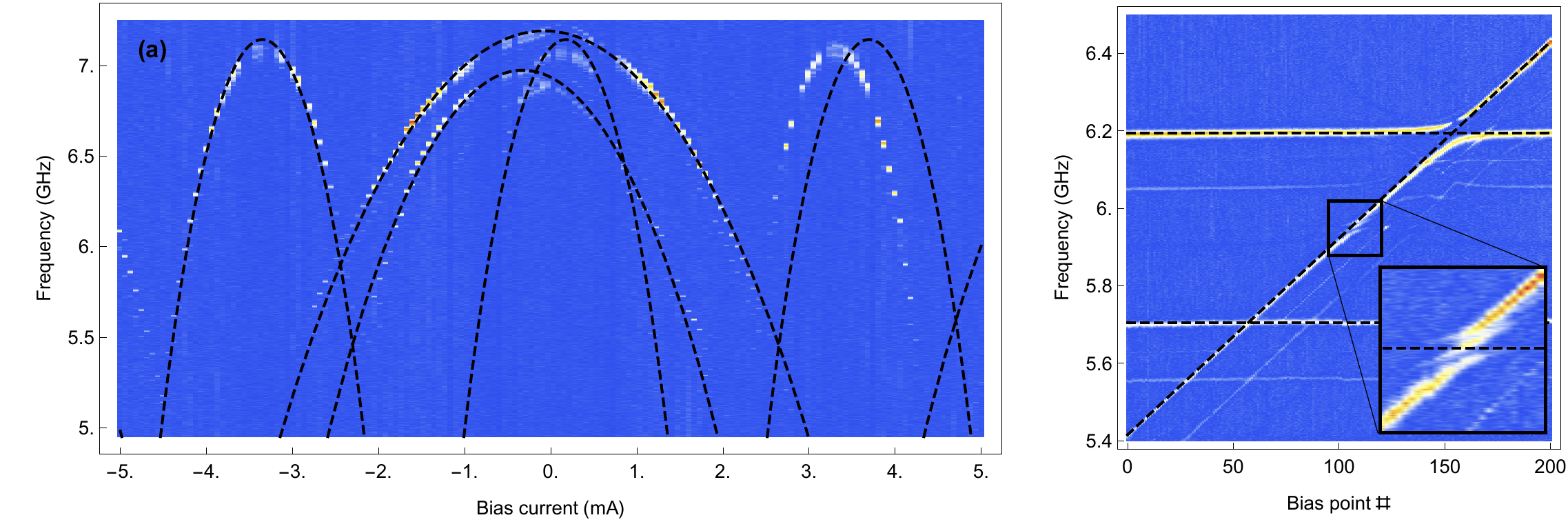}
		\caption{
			(a) Spectroscopy of the three transmon qubits. The x axis indicates the electrical current value driven through the larger magnetic coil mounted to the cavity. The y-axis shows the frequency of the excitation tone sent to the input of the cavity together with the measurement tone. The color indicates transmission of the measurement tone through the cavity at its resonance frequency, recalibrated at each bias current with no excitation. The background transmission was subtracted for each bias current for clarity.
			The dashed black curves indicate a fit to the expected magnetic dependencies of the transmon frequencies allowing extraction of the mutual inductances between the coil and SQUID loops for each qubit. 
			(b) Once the mutual inductance matrix between the coils and the qubits is extracted one can realize individual control of the qubit frequencies. In this example two of the qubits are kept at the chosen frequencies of $5.705\,$ and $6.195\,$GHz, respectively,  while the third qubit was made to change its frequency linearly with the bias parameters. The dashed black lines are shown as a guide for the eye. The inset shows a zoom in on the avoided level crossing observed around $(5.705 + 6.195)/2 = 5.950\,$GHz (the frequency is indicated with dashed line in the inset).  
			} 
		\label{fig1}
	\end{center}
\end{figure*}

To avoid losses in potential gaps in any mating faces, we evaporated $\sim 2\,\mu$m of Al on all Cu surfaces which are in contact with the Al cavity (the copper insert after evaporation of Al and Al parts are shown in Fig,~1(d)). The evaporation was completed with an e-beam evaporator by attaching the Cu insert to the evaporator's sample holder. The copper inserts surface facing the interior of the cavity was covered with Al tape and the evaporator sample holder was tilted at a $45^{\circ}$ angle and was constantly rotated about the deposition direction during evaporation. After that procedure, and sealing the cavities mating faces with indium wire to ensure light tightness, the cavity demonstrated a $Q = 102@000$ with an order of magnitude improvement compared to the quality factor of the bare copper cavity. An additional measurement of the same cavity several months later showed decrease of Q to the value of $80@000$ which we associated with oxidation of evaporated Al film. The quality factor of the cavity showed no power dependence within the wide range of powers from approx. -50$\,$dBm to -160$\,$dBm.  The fact that HFSS simulation predicted even higher quality factor suggests that further optimization to decrease internal losses is potentially possible.

To demonstrate individual control of three superconducting qubits with an external magnetic field we fabricated three transmon type qubits on an intrinsic Si substrate of size $10\,$mm by $5\,$mm, with a thickness of $500\,\mu$m (see Fig.~1(c)). All qubits were fabricated in a single step of electron beam lithography followed by shadow evaporation of two Al layers of 20$\,$nm and 40$\,$nm thick with an oxidation step between the depositions. Each qubit consists of two planar capacitor plates 700$\,\mu$m wide by $350\,\mu$m tall. The plates are separated by 50$\,\mu m$ and connected via a line interrupted by a micron size DC SQUID, playing the role of a magnetically tunable Josephson junction. The chip was placed in between the two halves of the cavity and magnetic flux supplied by three superconducting coils (two smaller and one larger) attached to the copper cavity was used to individually control the transition frequencies of all three qubits. The coils had 4000 turns of $25\,\mu$m superconducting wire and their diameters were $6\,$mm and $2\,$mm for large and small coils, respectively. The coil setup is virtually identical to one used to control three superconducting qubits coupled to a coplanar waveguide cavity on a chip~\cite{Fink2009}.

The input and output ports of the cavity were then coupled asymmetrically for measuring qubit spectroscopy in transmission, with corresponding external quality factors of $Q_{in} \simeq 50@000$ and $Q_{out} \simeq 4@000$ (the output coupling was substantially increased to also perform time domain characterization of the qubits, data is not presented here).  The fundamental mode of the cavity was shifted down to $7.295\,$GHz due to the insertion of the chip. A typical spectrum of all of the qubits as function of current applied to one of the coils is shown in Fig.~2(a). At each point in magnetic field, the resonant frequency of the cavity was found through a transmission measurement, where all of the qubits were in their ground states at given detunings from the cavity. Subsequently, the frequency of the measurement tone was set on resonance and the transmitted signal was amplified, down-converted and digitized to determine its amplitude and phase. Together with the measurement tone, an additional excitation tone was applied to the input of the cavity (see Fig.~1(a)). When the frequency of the excitation tone matched the frequency of one of the qubits, the transmission of the measurement tone changed due to a dispersive shift of the cavity frequency induced by the excitation of that particular qubit. As the distance between qubits is small compared to the size of the magnetic coils, any current applied through one coil effects all qubits, as can be seen in Fig~2(a). The mutual inductances between the coils and qubits' SQUID loops can be determined by fitting the spectrum to the expected theoretical function: $f_i = (E_{{\rm Jmax,i}}|\cos (\pi \Phi_i/\Phi_0)|E_c)^{1/2} - E_c$, where $\Phi_i = \Phi_{{\rm offset}, i} + \sum_j M_{ij} I_{{\rm period}, ij}$ and $\Phi_0$ is the magnetic flux quantum. The charging energy of the transmon is fixed by the geometry of capacitor pads and was chosen to be $E_c = 130\,$MHz. The maximum Josephson energy of the SQUID at zero magnetic field $E_{{\rm Jmax}, i}$, the flux offset $\Phi_{{\rm offset}, i}$ and the mutual inductances $M_{ij}$ are the determined by the fitting the modulation of the frequency of qubit $i$ when current is applied to coil $j$.
In the case that the coupling of the coils to different qubits are sufficiently different, the matrix can then be diagonalized, inverted and used for arbitrary control of the qubits frequencies. To demonstrate this, we fixed the frequencies of two qubits at $5.705\,$GHz and $6.195\,$GHz and tuned the frequency of the third qubit linearly (See Fig.~2(b)). Fig.~2(a) shows that the frequency modulations of the qubits are not perfectly periodic most probably from the rearrangement of pinned magnetic vortices when changing external magnetic flux. Nonetheless, with fine adjustments of the mutual inductance matrix it is always possible to achieve precise frequency control for all qubits.

This individual control over the qubits allowed us, for example, to see interesting spectrum features with clarity. In particular, an exchange interaction between the qubits mediated by resonator field photons manifested in the avoided level crossings of the qubit spectral lines~\cite{Majer2007} can be clearly seen for the qubit at $6.195\,$GHz. Within the avoided level crossing, the eigenstates of the system are in superpositions of the ground and excited states of the non-interaction qubits. The top spectral line of the avoided level crossing at $\sim 6.195\,$GHz shows a clear gap corresponding to the anti-symmetric dark state which cannot be excited by symmetry~\cite{Filipp2011b}. 

It is interesting to note that in addition to the well-known features of the spectrum described, there are also some spectrum particularities not described by the conventional models. More specifically while the avoided level crossing at $6.195\,$GHz is clearly visible, there is no observable avoided level crossing for another pair of qubits at $5.705\,$GHz. This phenomenon cannot be accounted by the Tavis-Cummings Hamiltonian, describing several qubits interacting to a single quantized mode of electromagnetic field~\cite{Tavis1968}. We speculate that the coupling through the resonator is compensated by the direct dipole-dipole interaction between the closest qubits which gives rise to the exchange interaction of opposite sign. Similar behavior has been observed but not reported for the system of three qubits~\cite{Baur2012} but has never been studied in detail. 

In addition, a smaller avoided level crossing of the third qubit was observed half-way between the frequencies of the first and second qubits at $\sim 5.950\,$GHz. From the frequency matching conditions it may be attributed the process where two excitations of the third qubit are exchanged with excitations of the first and second qubits. This process can not be explained neither by the Tavis-Cummings model nor by inclusion of the direct dipole-dipole coupling. There is also a more complex higher order transition pattern that can also be seen on the background of the spectrum as faint lines. These complex peculiarities of the spectrum may be subject of further investigation using 3D hybrid cavities with flux control.

In conclusion, this new type of hybrid 3D cavity permits us to reach quality factors an order of magnitude greater compared to pure normal metal cavities without loosing the feature of magnetic field tunability. This ability makes hybrid cavities an attractive choice for circuit QED experiments with flux tunable quantum systems in regimes where cavity losses are critical for performance.


%
%

%

\begin{acknowledgments}
We thank Prof. Y. Nakamura for providing us with a three-qubit sample. We also thank Dr. V. Monarkha for helping us with measurement of power dependence of the quality factor of the cavity.  M.J., A.F. were supported by the Australian Research Council Centre of Excellence CE110001013. Y.R. was supported by the Discovery Project DP150101033. AF was supported in part by the ARC Future Fellowship FT140100338. 
\end{acknowledgments}

\bibliography{Z:/RefDB/SQDRefDB}

\end{document}